%% file: main.tex
\documentclass{llncs}
\title{DistMS: A Non-Portfolio Distributed Solver\\for Maximum Satisfiability}
\pdfoutput=1
\usepackage[utf8]{inputenc}
\usepackage[T1]{fontenc}
\usepackage{graphicx}
\usepackage[ruled,linesnumbered]{algorithm2e}

\newtheorem{exmp}{Example}[section]
\newcommand{\DistMS}{{\tt DistMS} }

\begin{document}

\author{Miguel Neves \and In\^es Lynce \and Vasco Manquinho}
\authorrunning{Neves, Lynce and Manquinho} 
\tocauthor{Miguel Neves, Ines Lynce, Vasco Manquinho}
\institute{INESC-ID / Instituto Superior T\'ecnico, Universidade de Lisboa, Portugal\\
  \email{\{neves,ines,vmm\}@sat.inesc-id.pt}}

\maketitle

\begin{abstract}
  The most successful parallel SAT and MaxSAT solvers follow a
  portfolio approach, where each thread applies a different algorithm 
  (or the same algorithm configured differently) to solve a given problem
  instance.
  The main goal of building a portfolio is to diversify the search process 
  being carried out by each thread. As soon as one thread finishes, the
  instance can be deemed solved.
  In this paper we present a new open source distributed solver for 
  MaxSAT solving that addresses two issues commonly found in multicore 
  parallel solvers, namely memory contention and scalability.
  Preliminary results show that our non-portfolio distributed MaxSAT solver
  outperforms its sequential version and is able to solve more instances
  as the number of processes increases.
\end{abstract}

\input{prelim}

\input{solver}

\input{results}

\bibliography{main}
\bibliographystyle{splncs03}

\end{document}

%% file: prelim.tex
\section{Context and Motivation}
\label{sec:prelim}

The Maximum Satisfiability (MaxSAT) problem can be defined as an optimization version
of the Propositional Satisfiability (SAT) problem. Given an (usually unsatisfiable)
CNF formula $\phi$, the MaxSAT problem can be defined as finding an assignment
to problem variables such that it minimizes the number of unsatisfied clauses 
in $\phi$. In partial MaxSAT, given a CNF formula $\phi = \phi_S \cup \phi_H$,
the goal is to find an assignment such that it minimizes the number of
unsatisfied soft clauses in $\phi_S$ while satisfying all hard clauses in $\phi_H$.
Although there are weighted variants of MaxSAT~\cite{manya-handbook09}, 
in this paper we focus on partial MaxSAT.

Let $\phi_R$ denote the relaxation of a partial MaxSAT formula $\phi$.
In $\phi_R$ we associate a fresh relaxation variable $r_j$ with each 
soft clause $\omega_j$ in $\phi_S$ such that $\phi_R = \phi_H \cup \{(\omega_j \vee r_j) : \omega_j \in \phi_S \}$.
Notice that finding an assignment to the variables in $\phi_R$ such that 
it minimizes the number of relaxation variables assigned value 1 while 
satisfying all clauses is equivalent to solving the partial MaxSAT formula 
$\phi$. Hence, a common approach for solving partial MaxSAT is to relax
$\phi$ and iteratively call a SAT solver on $\phi_R$ with an
additional constraint $\sum r_j \le b$ encoded into 
CNF~\cite{DBLP:conf/sat/FuM06}.
Initially, we can define $b = |\phi_S|$. If a solution is found such 
that $\mu$ relaxation variables are assigned value 1, then $b$ is updated 
to $\mu-1$. The algorithm ends when the formula becomes
unsatisfiable and the optimal solution is the last one found with 
value $\mu$.

The described algorithm performs a linear search on the number of satisfiable
soft clauses. More recently, several algorithms have been proposed
that take advantage of the ability of SAT solvers to provide
an unsatisfiable sub-formula~\cite{DBLP:conf/date/ZhangM03}. 
These unsatisfiability-based algorithms have been shown very successful 
for solving industrial instances. We
refer the reader to the literature for details~\cite{morgado-constraints13}.

Due to the predominance of multicore architectures instead of higher 
frequency CPUs, recent work on MaxSAT solving has been deviating to 
the design of parallel solvers. The most successful parallel algorithms 
implement a portfolio of sequential solvers. The portfolio may include 
both different algorithms or the same algorithm with different configurations. 
In parallel solutions, all the cores of a typical computer access primary 
memory through the same BUS, and no two cores can use the BUS at the 
same time. Therefore, as the number of cores increases, so does 
contention on the access to primary memory, hindering the scalability 
of parallel algorithms. 

Another alternative to sequential algorithms are distributed algorithms. 
On one hand, these algorithms do not suffer from the drawback mentioned 
above. On the other hand, distributed algorithms are naturally designed 
to surpass the limitations on diversification of portfolio-based approaches. 
This paper describes {\tt DistMS}, a distributed MaxSAT solver that implements 
two distributed algorithms. The first algorithm splits the search space 
by assigning different upper bound values of the optimum solution to 
different processes. The second algorithm is based on choosing a subset 
of the problem’s variables and dividing the possible combinations of 
values for those variables among the processes.

%% file: solver.tex
\section{Distributed MaxSAT Solver \DistMS}
\label{sec:distms}

In this section we present the distributed algorithms for MaxSAT implemented
in the \DistMS solver.
The first algorithm is an adaptation of the parallel Search Space Splitting 
approach, first proposed by Martins et al. \cite{PWBO2,PWBO3}. The second 
algorithm is based on the guiding paths splitting strategy \cite{PSATO,LOOKAHEAD,Dolius}, 
which has been shown to be successful in parallel and distributed SAT solvers.

The architecture of \DistMS is composed of one master and multiple slave 
processes. The master process mediates the whole communication between the 
slave processes, being responsible for assigning tasks to slaves and handling  
communication. On the other hand, slaves wait for a task to be given by the 
master, process that task and send the result back to the master. 

Unlike other distributed solvers~\cite{Dolius}, slave processes in \DistMS do 
not communicate directly. All communication
is established between the master and the slave processes. The main goal is to 
minimize the changes in the slave behaviour such that any MaxSAT algorithm 
can be easily incorporated into the \DistMS solver. 
Although one might think that this would be a bottleneck in the master
process, such behaviour has not been observed, even when the number of
processes increases to several tens of slaves.

\subsection{Search Space Splitting Algorithm}
\label{section:sssalgorithm}

Given $n$ processes, the Search Space Splitting algorithm is composed by $1$ 
master, $1$ unsatisfiability-based process and $n-2$ linear search processes. The master process is responsible for keeping a lower bound $\lambda$ and an upper 
bound $\mu$ on the optimum solution of the MaxSAT instance. The lower bound is initially 0. 
The master process starts with a SAT call containing only the hard clauses $\phi_{H}$. 
Note that if this SAT call is unsatisfiable, then the solver terminates immediately, 
returning that the instance is not satisfiable. Otherwise, the number of 
unsatisfiable soft clauses provides an initial upper bound $\mu$.

The main goal of the search splitting algorithm is to split the set of possible values for the 
optimal solution. Given $k$ linear search slave processes, $p_{1}, \ldots, p_{k}$, the interval
defined by the lower bound $\lambda$ and upper bound $\mu$ is split across the $k$ processes. The initial bounds set is $\lbrace b_{0}, b_{1}, ..., b_{k-1}, b_{k} \rbrace$ with $b_0 = \lambda$ and $b_k = \mu - 1$ and with each process $p_i$ being responsible for checking if a given tentative bound $b_i$ is either a 
lower or upper bound on the optimal solution.

A slave process $p_i$ executes a SAT call on a relaxed MaxSAT formula $\phi_R$ with the additional 
constraint $\sum r_j \le b_i$ encoded into CNF~\cite{bailleux-cp03,qmaxsat-jsat12}.
If the formula given to process $p_i$ is unsatisfiable, then $b_i$ is a lower bound and 
$\lambda$ can be updated. Otherwise, $b_i$ is 
an upper bound and $\mu$ can be updated with the number of relaxation variables $r_j$ 
assigned value 1.

Initially, the value of $b_i$ for process $p_i$ is given by $\lfloor i \times \frac{\mu-1}{k} \rfloor$
since we have $\lambda = 0$. The master process maintains a sorted set $B$ of bounds to be
checked by the slaves.

\begin{exmp}
Let $\mu = 37$ be the initial upper bound. Given $k = 6$ linear search processes, the initial bounds set 
is $B = \lbrace 0, 6, 12, 18, 24, 30, 36 \rbrace$.
\end{exmp}

If a given slave process finds a new lower bound, then $\lambda$ is updated and
all values smaller than $\lambda$ are removed from the bound set $B$. Otherwise, if a slave
process finds a new upper bound, then $\mu$ is updated, all values larger than $\mu$ are 
removed and $\mu$ is added to $B$. 

Let $B = \lbrace b_{0}, b_{1}, ..., b_{k-1}, b_{k} \rbrace$ be the current bounds set.
If a given slave process $p_i$ needs a new bound to search on, then the master chooses a
pair $(b_{m-1}, b_{m})$ of contiguous values such that $b_{m} - b_{m-1} \ge b_{j} - b_{j-1}$ 
for all $1 < j \le k$. A new tentative bound $b_i = \frac{b_{m} + b_{m-1}}{2}$ is computed 
and $b_i$ is added to $B$. Furthermore, the new bound is sent to slave process $p_i$ as
a tentative bound.

\begin{exmp}
Let $B = \lbrace 5, 12, 22, 27, 40 \rbrace$ be a bounds set. Suppose that 
$p_{1}$ finds that $26$ is an upper bound. In this case, $B$ is updated to
$B = \lbrace 5, 12, 17, 22, 25 \rbrace$ where $17$ is the new tentative bound
for slave process $p_{1}$.
Next, if $p_{1}$ finds that $17$ is a lower bound, then the bounds set is 
updated to $B = \lbrace 18, 20, 22, 25 \rbrace$ where $20$ is $p_{1}$'s new tentative
bound.
\end{exmp}

Additionally to the slave processes that check tentative bounds, \DistMS also includes
a slave process executing an unsatisfiability-based algorithm. 
This is mainly to be able to quickly update the lower bound in few iterations, 
thus constraining the tentative bounds to be provided to other slave processes
by the master.

An optimal solution is found when the lower bound $\lambda$ is equal to the upper 
bound $\mu$. When this occurs, the master process aborts the execution of the 
remaining processes and terminates, returning $\mu$ as the optimum value.

\subsection{Guiding Paths with Lookahead Algorithm}
\label{section:gpalgorithm}

Heule et al. \cite{LOOKAHEAD} already proposed a parallel SAT algorithm that 
initially uses a \textit{lookahead} solver to generate guiding
paths in order to split the search tree. Lookahead solvers apply sophisticated reasoning 
at each branching step in order to guide the search more effectively. The algorithm 
described throughout the rest of this section is an extension of the previous 
approach to distributed MaxSAT. 

Given $n$ processes, the guiding path algorithm in \DistMS is composed by $1$ master, $n - 2$ guiding path solver processes and $1$ linear search process. The master starts by generating a queue of guiding paths to be solved by the slave 
processes while waiting for an initial upper bound $\mu$ from the linear search process. 
The guiding paths are heuristically sorted and given to available
slave processes with the best upper bound computed thus far. Each slave applies a
linear search MaxSAT algorithm~\cite{QMAXSAT} and returns the best solution found
for the given path to the master. If the newly found solution improves on the
previous one, it is saved and the upper bound $\mu$ is updated.

Note that unlike other guiding path solver architectures, 
the number of initial guiding paths is usually much larger than the number of 
slaves.
Hence, when a given guiding path is solved, the master immediately removes the
first guiding path from the queue and sends it to the slave. The MaxSAT
instance is considered solved when the guiding path queue becomes empty.

When the master sends a guiding path to a slave process, it also provides
the current upper bound $\mu$. The working formula on the slave process
contains a relaxation of the MaxSAT formula $\phi_R$ and a cardinality
constraint $\sum r_j \le \mu-1$ encoded into CNF. The guiding path literals
are considered assumptions in the SAT solver calls occurring in the slave
process. Therefore, if the working formula is unsatisfiable, the slave
process is able to provide a reason for the unsatisfiability of the
formula to the master. When the unsatisfiability does not depend on
the guiding path, one can conclude that the working formula is
not satisfiable due to the cardinality constraint and 
$\mu$ is a lower bound of the MaxSAT formula. 
As a result, the previously found solution $\mu$ is optimal and the
solver can terminate, even if there are guiding paths in the queue.

\begin{algorithm}[t]

\SetAlgoLined
\SetKwFunction{GenerateGuidingPaths}{GenerateGuidingPaths}
\SetKwFunction{IncrementCutoff}{IncrementCutoff}
\SetKwFunction{DecrementCutoff}{DecrementCutoff}
\SetKwFunction{Propagate}{Propagate}
\SetKwFunction{AnalyzeAndLearn}{AnalyzeAndLearn}
\SetKwFunction{Vars}{Vars}
\SetKwFunction{ChooseVariable}{ChooseVariable}
\SetKwFunction{ChoosePolarity}{ChoosePolarity}

\SetKwProg{procedure}{Procedure}{}{}

\procedure{\GenerateGuidingPaths{$\phi$, $C$, $D$, $I$, $\theta$}}
{

\IncrementCutoff{$\theta$}

$(\phi, I) \leftarrow$ \Propagate{$\phi$, $D$, $I$}

\If{$\phi$ is unsatisfied by $D \cup I$ or $\left| D \right| + \log_{2}{\left| \phi \right|} > 25$}
{
\DecrementCutoff{$\theta$}
}

\If{$\phi$ is unsatisfied by $D \cup I$}
{
\AnalyzeAndLearn{$\phi$, $D$, $I$}

\KwRet{$C$}
}

\If{$\left| D \right| \times \left| D \cup I \right| > \theta \times |$ \Vars{$\phi$} $|$}
{
\KwRet{$C \cup \lbrace D \rbrace$}
}

$x \leftarrow $ \ChooseVariable{$\phi$, $D$, $I$}

$l \leftarrow $ \ChoosePolarity{$\phi$, $x$}

$C \leftarrow $ \GenerateGuidingPaths{$\phi$, $C$, $D \cup \lbrace l \rbrace$, $I$, $\theta$}

\KwRet{\GenerateGuidingPaths{$\phi$, $C$, $D \cup \lbrace \neg l \rbrace$, $I$, $\theta$}}

}

\caption{Guiding Path generation algorithm \cite{LOOKAHEAD}}\label{alg:gp-generation}

\end{algorithm}

The pseudo-code for the guiding path generation procedure is presented in algorithm \ref{alg:gp-generation}. This procedure receives as input a CNF formula $\phi$, the set $C$ of guiding paths computed so far by the procedure, the current partial assignment $D$, the set $I$ of literals that are implied by the partial assignment $D$ and a cutoff value $\theta$. Formula $\phi$ corresponds to the hard clauses of the MaxSAT instance.

The algorithm starts by incrementing the cutoff value $\theta$ (line 2). This is done to prevent $\theta$ from being reduced too much by the decrement rule in line 4 and as a consequence generating too small guiding paths. The initial cutoff value is $1000$ as specified by Heule et al. \cite{LOOKAHEAD}. In practice, $\theta$ is incremented by $5 \%$. Next, unit propagation is applied to simplify $\phi$ and update set $I$ (line 3). The algorithm then checks if $\phi$ is unsatisfied by the current assignments (line 4). In this case, $\theta$ is decremented (line 5). $\theta$ is also decremented if $\left| D \right| + \log_{2}{\left| \phi \right|} > 25$ (line 4). This rule prevents the guiding path generation process of going too deep in the search tree and generating too many guiding paths. In practice, $\theta$ is decremented by $30 \%$.

If $\phi$ is unsatisfied, then the procedure applies conflict analysis \cite{CLAUSELEARNING1,CLAUSELEARNING2} and learns a new clause (line 8), similarly to a CDCL SAT solver \cite{CDCLSAT}. This may prevent the procedure from generating guiding paths that unsatisfy $\phi$. If $\phi$ is not unsatisfied, then the algorithm checks if the cutoff has been triggered (line 11). If so, $D$ is returned as a guiding path. The cutoff condition takes into account the number of branching steps and the total number of assignments, explicit and implied, in the current node of the search tree.

If the cutoff is not triggered, then an unassigned variable $x$ is chosen heuristically to be added to $D$ (line 14). Given a variable $x$, we denote as $eval_{cls}(x)$ ($eval_{cls}(\neg x)$) the sum of the weights of the clauses that are reduced by the assignment $x = 1$ ($x = 0$) but are not satisfied. The clauses are weighted in a way such that a clause with length $k$ has a weight five times larger than a clause with length $k + 1$. Variables are ranked by $eval_{cls}(x) \times eval_{cls}(\neg x)$ and ties are broken by $eval_{cls}(x) + eval_{cls}(\neg x)$.

\begin{exmp}
Let $\phi = \lbrace (x_{1} \vee x_{2} \vee x_{3}), (x_{2} \vee \neg x_{3}), (\neg x_{1} \vee x_{2}) \rbrace$ be a CNF formula. Hence, $eval_{cls}(\neg x_{3}) = 6$ (clauses with length $2$ and $3$ have weights $5$ and $1$, respectively, if $3$ is the maximum clause size).
\end{exmp}

In practice, a variation of the $eval_{cls}$ heuristic, referred to as $eval_{wl}$, is used to rank variables. Given a literal $l$, the only difference is that instead of considering all the clauses in $\phi$, only the clauses {\em watching} \cite{CHAFF} literal $l$ are considered in the computation of $eval_{wl}(l)$. Also, only variables in soft clauses are considered when choosing a new variable.

After choosing a variable $x$, another heuristic is used to decide which truth value will be tested first (line 15). We choose the direction based on the number of clauses that will be unsatisfied after assigning $x$. The branch to be explored first is the one that unsatisfies a smaller number of soft clauses. Ties are broken choosing the direction that satisfies more soft clauses. The rationale for this is that the branch that unsatisfies less soft clauses is more likely to reach an upper bound closer to the optimum value. Algorithm \ref{alg:gp-generation} is then repeated for $x$ and $\neg x$ (lines 16 and 17).

Note that, since the master sorts guiding paths as they are generated, in the long run the sorting heuristic dominates the polarity heuristic. The priority is given to the least restricting guiding path and ties are broken by choosing the one that was generated first.

If there are idle processes and there are no guiding paths left in the queue, then one of the paths currently being solved is chosen to be further split into new guiding paths. We choose the path $g$ that was assigned first. Algorithm \ref{alg:gp-generation} is re-invoked, but this time with $D = g$ and $I$ updated accordingly. When algorithm \ref{alg:gp-generation} is re-invoked with $D \neq \emptyset$, $\theta$ is initialized as $5000$, or else too few guiding paths would be generated.

%% file: results.tex
\section{Experimental Results and Discussion}

The results in Table~\ref{table:results} were obtained on the partial 
MaxSAT crafted and industrial instances of the MaxSAT evaluation of 2013. 
{\tt DistMS} was implemented on top of OpenWBO~\cite{DBLP:conf/sat/MartinsML14}
and different configurations of {\tt DistMS} are compared against the sequential
counterparts. {\bf MSU3} refers to the OpenWBO's unsatisfiability-based 
MSU3 algorithm, {\bf LinearSU} to the OpenWBO's linear search algorithm, 
{\bf GP-$n$:$m$} to guiding paths with $n$ processes per each one of the $m$ 
machines and {\bf SSS-$n$:$m$} to search space splitting with $n$ processes 
per each one of the $m$ machines.
For each instance, algorithms were executed with a timeout of 1800 seconds
(wall clock time) and a memory limit of 4 GB per process. The tests were conducted on 
a cluster of machines with 4 AMD Opteron 6376 (2.3 GHz) and 128 GB of RAM, 
running Debian jessie. 

\begin{table*}[t]
  \centering
  \footnotesize
  \caption{Experimental evaluation of {\tt DistMS}}
  \begin{tabular} { c || r | r | r | r | r | r | r }
    \textbf{Instance Group} & \textbf{Total} & {\bf MSU3} & {\bf LinearSU} & \textbf{GP-2:4} & \textbf{GP-2:8} & \textbf{SSS-2:4} & \textbf{SSS-2:8} \\
    \hline
    crafted & 377 & 260 & 283 & 290 & 294 & 279 & 278 \\
    industrial & 627 & 551 & 524 & 518 & 515 & 562 & 564 \\
  \end{tabular}
  \label{table:results}
\end{table*}

Experimental results show that search space splitting (SSS) performs better
in industrial instances, while using guiding paths (GP) allows {\tt DistMS} to
perform better in crafted instances. SSS solves more instances than the sequential
solver and slightly increases its performance with a growing number of processes.
However, gains are small, since the solver quickly converges to near the 
optimum bound, and then the diversification of the search is small.

The GP approach fails to perform in industrial instances. Unlike the SSS, in
the GP approach it is hard to converge to the optimum in industrial instances.
Nevertheless, observe that the GP approach is the best performing in crafted
instances and it continues to improve as the number of processes grows.

This paper proposes the first distributed MaxSAT solver. Although previous
multicore parallel approaches have been proposed, they fail to scale when the
number of threads increases beyond 8 threads, since these are based in a
portfolio of sequential solvers. As a result, {\tt DistMS} integrates two
non-portfolio strategies, namely search space splitting on the number of
unsatisfied soft clauses and generation of guiding paths. Experimental
results show that we are able to improve on the sequential solvers, but
the scalability is still unclear. Given the mixed results from both
approaches in different sets of instances, as future work we propose to 
integrate them into a unifying framework for distributed MaxSAT solving.